\documentclass[12pt]{iopart}
\usepackage{iopams,graphicx}
\begin{document}

\title{WMAP 5-year constraints on lepton asymmetry and radiation energy density: Implications for {\sc Planck}}

\author{L.A. Popa and  A.Vasile}{
\address{ISS Institute for Space Sciences Bucharest-Magurele, Ro-077125 Romania}

\eads{\mailto{lpopa@venus.nipne.ro},
\mailto{avasile@venus.nipne.ro}}

\date{\today}

\begin{abstract}

In this paper we set bounds on the radiation content of the Universe and neutrino properties by using the WMAP-5 year CMB measurements complemented  with 
most of the existing CMB  and  LSS data (WMAP5+All),
imposing also self-consistent BBN constraints on the primordial helium abundance.\\
We consider lepton asymmetric cosmological models parametrized by the neutrino degeneracy parameter $\xi_{\nu}$ and the variation of the relativistic degrees of freedom, $\Delta  N^{oth}_{eff}$, due to possible other physical processes  occurred between BBN and  structure formation epochs.

We get a mean value of the effective number of relativistic neutrino species of $N_{eff}=3.256^{+0.607}_{-0.641}$ (68\% CL),  bringing an important improvement over the similar result obtained from WMAP5+BAO+SN+HST data \cite{Komatsu}.\\
We also find a strong correlation between $\Omega_mh^2$ and $z_{eq}$, showing that we observe  $N_{eff}$ mainly via the effect of $z_{eq}$, rather than via  neutrino anisotropic stress as claimed by the WMAP team  \cite{Komatsu}.
 
WMAP5+All data provides a strong bound on helium mass fraction of $Y_p=0.2486 \pm 0.0085$ (68\% CL), that rivals the  bound on $Y_p$ obtained from the conservative analysis of the present data on helium abundance.

For neutrino degeneracy parameter  we find a bound of  $-0.216 \leq \xi_{\nu} \leq 0.226$ (68\%CL), that represent an important improvement over the similar result obtained by using the WMAP 3-year data.
 
The inclusion in the analysis of LSS data reduces  the upper limit 
of the neutrino mass to $m_{\nu} < 0.419$ eV (95\% CL) with respect to the values obtained from the analysis from WMAP5-only data \cite{Dunkley} and WMAP5+BAO+SN+HST data \cite{Komatsu}.

We forecast that the CMB temperature and polarization measurements observed 
with high angular resolutions and sensitivities by the future {\sc Planck} satellite will reduces the errors on $\xi_{\nu}$ and $Y_p$ down 
to $\sigma(\xi_{\nu}) \simeq 0.089$ (68\% CL) and $\sigma(Y_p)=0.013$ (68\% CL) respectively, values fully consistent with the BBN bounds on these parameters.

This work has been done on behalf of {\sc Planck}-LFI activities.

\end{abstract}
\pacs{CMBR theory, dark matter, cosmological neutrinos, big bang nucleosynthesis  }

\maketitle

\section{Introduction}

The radiation budget of the Universe relies on a strong theoretical prejudice:
apart from the Cosmic Microwave Background (CMB) photons, the relativistic background
would consist of neutrinos and of possible contributions from other relativistic relicts.
The main constraints on the radiation energy density come either from the very early Universe, where the radiation was the dominant source of energy, or from the observation of cosmological perturbations which
carry the information about the time equality between matter and radiation.

\vspace{0.3cm}
In particular, the primordial light element abundance predictions in the standard theory
of the Big Bang Nucleosynthesis (BBN) \cite{Wagoner,Olive,Burles,Eidelman}
depend on the baryon-to-photon ratio, $\eta_B$, and the
radiation energy density at the BBN epoch (energy density of order MeV$^4$), usually parametrized by the effective number of relativistic neutrino species, $N_{eff}$.

Meanwhile, the number of active neutrino flavors have been fixed by  $Z^0$ boson decay
width to $N_{\nu}=2.944 \pm 0.012$ \cite{Eidelman} and the combined study of the incomplete neutrino decoupling and the QED corrections indicate that the number of relativistic neutrino
species is $N_{eff}=3.046$ \cite{Mangano02}.  Any departure of  $N_{eff}$ from this last value would be due to non-standard neutrino features or to the contribution of other relativistic relics. \\
The solar and atmospheric neutrino oscillation experiments  \cite{Fuk98,Amb98} indicate the existence of non-zero neutrino masses in eV range.\\
There are also indications of neutrino oscillations with  larger 
mass-squared difference,
coming from the short base-line oscillation experiments \cite{Atha,Miniboone},
 that can be explained by adding one or two sterile neutrinos with eV-scale mass to the standard scheme with three active neutrino flavors (see Ref.\cite{Maltoni} for a recent analysis). Such results have impact on cosmology because sterile neutrinos can contribute to the number of relativistic degrees of freedom at the Big Bang Nucleosynthesis \cite{Cirelli}.
These models are subject to strong bounds on the sum of active neutrino masses from the combination of various cosmological data sets \cite{Rafelt,Find},
ruling out a thermalized sterile neutrino component with eV mass \cite{Seljak07,Mel}.\\
However, there is the possibility to accommodate the cosmological observations
with data from short base-line neutrino oscillation experiments by postulating
the existence of a sterile neutrino with the mass of few keV having a phase-space distribution significantly suppressed relative to the thermal distribution.\\
For both, non-resonant zero lepton number production and enhanced resonant production
with initial cosmological lepton number, the keV mass sterile neutrino  produced via small mixing angle oscillation conversion of thermal active neutrinos \cite{Aba02} provides  a
valuable Dark Matter (DM) candidate,
alleviating the accumulating contradiction between the $\Lambda$CDM model
predictions on small scales and observations by smearing out the small scale structure 
 \cite{DodWidr,Dol,Aba01,Sha06}.   \\
A recent analysis of X-ray and Lyman-${\alpha}$ data  indicate that keV sterile neutrinos
can be considered  valuable DM candidates only if
they are produced via resonant oscillations with non-zero
lepton number or via other (non-oscillatory) production
mechanisms \cite{Palazzo}.\\
On the other hand, the possible existence of new particles such as axions and gravitons, the time variation of the physical constants and  other non-standard scenarios (see e.g. \cite{Sarkar96} and references therein) could contribute to the radiation energy density at BBN epoch.

At the same time, more phenomenological extensions to the standard neutrino sector
have been studied, the most natural being the consideration of the leptonic asymmetry
\cite{Freese83,Ruffini83,Ruffini88}, parametrized by the neutrino degeneracy parameter $\xi_{\nu}=\mu_{\nu}/T_{\nu_0}$
[$\mu_{\nu}$ is the neutrino chemical potential and
$T_{\nu_0}$  is the present temperature of the neutrino background,
$T_{\nu_0}/T_{\rm cmb}=(4/11)^{1/3}$]. \\
Although  the standard model of particle physics predicts the value of leptonic asymmetry  of the same order as the value of the  baryonic asymmetry, $B \sim 10^{-10}$, there are many particle physics scenarios in which a leptonic asymmetry much larger can be generated \cite{Smith06,Cirelli06}.
One of the cosmological implications of a larger leptonic asymmetry is
the possibility to generate small baryonic asymmetry of the Universe through
the non-perturbative (sphaleron) processes \cite{Kuzmin85,Falcone01,Buch04}.
Therefore, distinguishing between a vanishing and non-vanishing $\xi_{\nu}$ at the BBN epoch is a crucial test of the standard assumption that sphaleron effects equilibrate the
cosmic lepton and baryon asymmetries.\\
The measured neutrino mixing parameters implies that neutrinos
reach the chemical equilibrium before BBN \cite{Dolgov02,Wong02,Beacom02}, so that  all
neutrino flavors are characterized by the same
degeneracy parameter, $\xi_{\nu}$, at this epoch.
The most important impact of the leptonic asymmetry on BBN is
the shift of the beta equilibrium between protons and neutrons and the increase
of the radiation energy density parametrized by:
\begin{eqnarray}
\Delta N_{ eff}(\xi_{\nu})=3\left[ \frac{30}{7}
\left(\frac{\xi_{\nu}}{\pi} \right)^2
+\frac{15}{7}\left( \frac{\xi_{\nu}}{\pi} \right)^4 \right] \,.
\end{eqnarray}

The BBN constraints on N$_{eff}$ have been recently reanalyzed
by comparing  the theoretical predictions and experimental data on the
primordial abundances of light elements, by
using the baryon abundance derived from  the
WMAP~3-year (WMAP3) CMB temperature and  polarization measurements
\cite{Spergel07,Nolta,Page}: $\eta_B=6.14 \times 10^{-10}(1.00\pm 0.04)$.
In particular, the $^4$He abundance, $Y_p$,  is quite sensitive to the value of $N_{eff}$.
In the analysis of Ref. \cite{Mangano07}, the conservative error of  helium abundance,
$Y_P=0.249 \pm 0.009$ \cite{Olive04}, yielded to $N_{eff}=3.1^{+1.4}_{-1.2}$ (95\% CL) in good agreement with the standard value,
but still leaving some room for non-standard values, while
more stringent error bars of helium abundance , $Y_p=0.2516 \pm 0.0011$ \cite{Izotov07}, leaded to $N_{eff}=3.32^{+0.23}_{-0.24}$ (95\% CL) \cite{Ichikawa07}.

The stronger constraints on the degeneracy parameter obtained from BBN  \cite{Serpico05} gives $-0.04 < \xi< 0.07$ (69\% CL), adopting  the conservative error analysis of $Y_p$ of Ref. \cite{Olive04} and $\xi=0.024 \pm 0.0092$ (68\% CL),
adopting the more stringent error bars of $Y_p$ of Ref. \cite{Izotov03}.

\vspace{0.2cm}
The CMB anisotropies and LSS matter density fluctuations power spectra carry the signature
of the energy density of the Universe at the time of matter-radiation equality
(energy density of order eV$^4$), making possible the measurement of $N_{eff}$ through its effects on the growth of cosmological perturbations. \\
The number of relativistic neutrino species influences the CMB power spectrum by changing  the time of matter-radiation equality that enhances the integrated Sachs-Wolfe effect, leading to a higher first acoustic Doppler peak amplitude. Also, the temperature anisotropy of the neutrino background 
(the anisotropic stress) acts as an additional source term for the gravitational potential \cite{Hu95,Trotta05}, changing the CMB anisotropy power spectrum at the level of $\sim 20$\%.\\    
The delay of the epoch of matter-radiation equality shifts the LSS matter power spectrum turnover position toward larger angular scales, suppressing the power at small scales. In particular, the non-zero neutrino chemical potential leads to changes in neutrino free-streaming length and neutrino Jeans mass due to the increase of the neutrino velocity dispersion \cite{Lattanzi05,Ichiki07}.

After WMAP3 data release there are many works aiming to constrain $N_{eff}$ from cosmological observations \cite{Seljak07,Spergel07,Mangano07,Han06,Cirelli06b,Kawa07}.
Their results suggest large values for $N_{eff}$ within 95\% CL interval, some of them  not including the standard value $3.046$ \cite{Seljak07,Spergel07,Mangano07}. Recently Ref. \cite{Han07} argues that the discrepancies are due to the treatment of the  scale-dependent biasing in the galaxy power spectrum inferred from the main galaxy sample of the Sloan Digital Sky Survey data release 2 (SDSS-DR2)  \cite{Teg04a,Teg04b}
and the large fluctuation amplitude reconstructed from  the Lyman$-\alpha$ forest data
\cite{McDonald05} relative to that inferred from WMAP3. \\
Discrepancies between  BBN  and cosmological data results on $N_{eff}$ was interpreted as 
evidence of the fact that further relativistic species are produced by particles decay between BBN and structure formation \cite{Cirelli06b,Kawa07}.
Other theoretical scenarios include the violation of the spin-statistics in the neutrino sector \cite{Dolgov05}, the possibility of an extra interaction between
the dark energy and radiation or dark matter, the existence of a Brans-Dicke field which could mimic the effect of adding extra relativistic energy density between BBN and structure formation epochs \cite{deFelice}. 

A lower limit to $N_{eff} > 2.3$ (95\% CL) was recently obtained from the analysis of the 
WMAP 5-year (WMAP5) data alone \cite{Dunkley}, while the combination of the WMAP5 data with 
distance information from baryonic acoustic oscillations (BAO), supernovae (SN) and Hubble constant measured by Hubble Space Telescope (HST), leaded to
$N_{eff}=4.4 \pm 1.5$ (68\% CL), fully consistent with the standard value \cite{Komatsu}.          

\vspace{0.3cm}
The extra energy density can be splitted in two distinct uncorrelated contributions, first due to net lepton asymmetry of the neutrino background and second due to the extra contributions from other unknown processes:
\begin{equation}
\Delta N_{eff}=\Delta N_{eff}(\xi)+ \Delta N^{oth}_{eff}\, .
\end{equation}
The aim of this paper is to obtain bounds on the neutrino lepton asymmetry and on the extra radiation energy density by using WMAP5 data in combination with most of the existing CMB and 
LSS measurements and self-consistent BBN priors on $Y_p$. 
We also compute the sensitivity of the future {\sc Planck} experiment \cite{Blue} 
for these parameters testing the restrictions on cosmological models with extra relativistic degrees of freedom expected from high precision CMB temperature and polarization anisotropy measurements.    

\section{Leptonic asymmetric cosmological models}

The density perturbations in leptonic asymmetric cosmological models
have been discussed in literature \cite{Kinney99,Crotty03,Lattanzi05,Ichiki07,Pastor99,Orito02}.
We applied them to modify the Boltzmann Code for Anisotropies in the Microwave Background (CAMB) \cite{Hu00,Cha,Lewis} to compute the  CMB temperature and polarization anisotropies
power spectra  and LSS matter density fluctuations power spectra for the case of
three degenerated neutrinos/antineutrinos with the total mass $m_{\nu}$ and degeneracy parameter $\xi_{\nu}$. As neutrinos reach their approximate chemical potential
equilibrium before BBN epoch \cite{Dolgov02,Wong02,Beacom02}, we consider in our computation that all three flavors of neutrinos/antineutrinos have the same degeneracy parameter $\xi_{\nu}$. \\
When the Universe was hot enough, neutrinos and antineutrinos of each flavor $\nu_i$ 
behave like relativistic particles with Fermi-Dirac phase space distributions:
\begin{eqnarray}
{\cal F}_{\nu_i}(q)=\frac{1}{e^{E_{\nu_i}/T_{\nu}-\xi_{\nu}}+1} \, ,
\hspace{0.2cm}
{\cal F}_{{\bar \nu_i}}(q)=\frac{1}{e^{E_{{\bar \nu_i}}/T_{\nu}-\xi_{{\bar \nu}}}+1} \,
\hspace{0.3cm} (i= e,\mu,\tau)
\end{eqnarray}
where $E_{\nu_i}=\sqrt{q^2+a^2m_{\nu_i}}$ is one flavor neutrino/antineutrino energy
and $q=ap$ is the comoving momentum.  Hereafter, $a$ is the cosmological scale factor ($a_0=1$ today). The mean energy density and pressure of one flavor of massive degenerated neutrinos and antineutrinos can be written as:
\begin{eqnarray}
\rho_{\nu_i}+\rho_{\bar \nu_i}=(k_BT_{\nu})^4\int^{\infty}_0 \frac{d^3q}{(2\pi)^3} \, q^2 E_{\nu_i}({\cal F}_{\nu_i}(q)+{\cal F}_{\bar \nu_i}(q))\, ,\\
3(P_{\nu_i}+P_{\bar \nu_i})=(k_BT_{\nu})^4\int^{\infty}_0 \frac{d^3q}{(2\pi)^3} \, \frac{q^2} {E_{\nu_i}}({\cal F}_{\nu_i}(q)+{\cal F}_{\bar \nu_i}(q))\, .
\end{eqnarray}
We modify in CAMB the expressions for the energy density and the pressure in the relativistic and non-relativistic limits for the degenerate case \cite{Ichiki07}
and  follow the standard procedure to compute the perturbed quantities  by expanding the phase space distribution function of neutrinos and antineutrinos into homogeneous  and perturbed inhomogeneous components \cite{Lewis,Ma95,Seljak96}.
Since the gravitational source term in the Boltzmann equation is proportional to the logarithmic derivative of the neutrino distribution function with respect to comoving momentum, ${\rm d} \ln ({\cal F}_{\nu_i}+{\cal F}_{\bar \nu_i})/ {\rm d} \ln q$, we also modify this term
to account for $\xi_{\nu} \neq 0$ \cite{Ichiki07,Pastor99}.

As mentioned before, the BBN theory gives strong constraints on $N_{eff}$ and $\xi_{\nu}$
by comparing the measured light element abundance with the theoretical predictions. The only free parameter is the baryon to photon ratio, $\eta_B=n_b/n_{\gamma}$, that is obtained from the determination of $\Omega_b h^2$ from CMB measurements.\\
In particular, the $^{4}$He mass fraction, $Y_p$, affects the CMB angular power spectra through its impact on different evolution phases of the ionization/recombination history \cite{Trota03}. \\
As previously demonstrated \cite{Iki_He,Hamann07,Ichikawa071}, the impact of self-consistent BBN prior on  $Y_p$  has a net impact on  parameter inference, improving the bounds on cosmological parameters compared to the analysis which treat $Y_p$ as a constant or free parameter. \\
We use the public BBN code PArthENoPE \cite{Pisanti} to compute the dependence of $Y_p$ on
$\Omega_bh^2$, $\Delta N_{eff}$ as given in equation (2) and $\xi_{\nu}$.   
The accuracy on $Y_p$ obtained by using PArthENoPE code is $\sim 10^{-4}$, being  only limited by the experimental uncertainty on the neutron lifetime \cite{Serpico04}.
We also modify the recombination routine Recfast v1.4 \cite{Scott99} of the CAMB code to account for  $Y_p$ dependence on $\Omega_bh^2$, $\Delta N_{eff}$ and $\xi_{\nu}$.\\
In our computation we implicitly assume that the value of $N_{eff}$ is not changed between the epoch of BBN and  matter-radiation equality.   
 
\section{Analysis}

We use the {\sc CosmoMC}  Monte Carlo Markov Chain (MCMC) public package \cite{Bridle}
modified for our extended $6+3$ parameter space to sample from the posterior distribution giving the following experimental datasets:
\begin{itemize}
\item The WMAP5 temperature and polarization CMB measurements \cite{Dunkley,Komatsu,Nolta08} complemented with the CMB measurements from Boomerang \cite{Netterfield02,Tavish05}, ACBAR \cite{Reichardt} and  CBI \cite{Readhead04} experiments.\\
\item The LSS power spectrum of the matter density fluctuations inferred from
the galaxy clustering data of the Sloan digital Sky Survey (SDSS) \cite{Teg04a,Teg04b,Tegmark06,Percival07} and Two-degree Field Galaxy Redshift Survey (2dFGRS) \cite{Cole05}. 
In particular, the luminous red galaxies (LRG) sample from  SDSS
data release 5 (SDSS-DR5) \cite{Tegmark06,Percival07} has more statistical significance  
than the spectrum retrieved from the SDSS main galaxy sample from data release 2 (SDSS-DR2)
\cite{Teg04a,Teg04b}  eliminating the existing tension between the power spectra
from SDSS-DR2 and 2dFGRS.
For this reason we include in our analysis the matter power spectra from SDSS-LRG and 2dFGRS. 
We consider SDSS-LRG data up to $k_{max} \simeq 0.2$h Mpc$^{-1}$ and the 2dFGRS data up to $k_{max} \simeq 0.14$h Mpc$^{-1}$ and apply the corrections due to the non-linearity behavior and scale dependent bias \cite{Tegmark06}, connecting the linear matter power spectrum, $P_{lin}(k)$,  
and the galaxy power spectrum, $P_{gal}(k)$, through:
\begin{equation}
P_{gal}(k)=b^2 \frac{1+Q_{nl}k^2}{1+1.4k}P_{lin}(k) \,,
\end{equation}
where the free parameters  $b$ and $Q_{nl}$ are marginalized.\\
\item The luminosity distance measurements of distant Type Ia Supernovae
(SNIa) obtained by Supernova Legacy Survey (SNLS) \cite{Astier06}
and the Hubble Space Telescope \cite{Riess04}.
\item The BBN constraints on $Y_p$ as obtained from PArthENoPE code, allowing $\Omega_b h^2$
$\Delta N_{eff}$ and $\xi_{\nu}$ to span the ranges indicated in Table 1.
\end{itemize}
Hereafter, we will denote  WMAP5+SDSS-DR5+2dFGRS+SNIa+BBN data set as WMAP5+All.

We perform our analysis in the framework of the extended
$\Lambda$CDM cosmological model described by $6+3$ free parameters:
\begin{eqnarray}
\Theta=( \underbrace{\Omega_b h^2,\Omega_{dm} h^2, H_0, z_{re}, n_s, A_s,}_{standard} \Omega_{\nu}h^2,
\xi_{\nu}, \Delta N^{oth}_{eff} )\, .
\end{eqnarray}
Here $\Omega_{b} h^2$ and $\Omega_{dm}h^2$ are the baryon and cold dark matter energy density parameters, 
$H_0$ is the Hubble expansion rate,
$z_{re}$ is the redshift of reionization, 
$n_s$ is the scalar spectral index of the primordial density perturbation power spectrum and 
$A_s$ is its amplitude at the pivot scale $k_*=0.002$~hMpc$^{-1}$. 
The additional three parameters denote the neutrino energy density  $\Omega_{\nu}h^2$, 
the neutrino degeneracy parameter $\xi_{\nu}$ and the contribution of extra relativistic degrees of 
freedom from other unknown processes $\Delta N_{oth}^{eff}$.
Table~1 presents the  parameters of our model, their fiducial values used
to generate the {\sc Planck}-like simulated power spectra and the prior ranges
adopted in the analysis.

\begin{table}
\begin{center}
\caption{The free parameters of our model, their fiducial values used
to generate the {\sc Planck}-like simulated power spectra and the prior ranges
adopted in the analysis.}
\end{center}
\vspace{0.3cm}
\begin{center}
\begin{tabular}{llll}
\hline \hline
 Parameter & Symbol &Fiducial value  & Prior range   \\
\hline \hline
Baryon  density&$\Omega_{b}h^2$ & 0.022 & 0.005 $\rightarrow$ 0.04 \\
Dark matter  density &$\Omega_{dm}h^2$ & 0.11& 0.01 $\rightarrow$ 0.5 \\
Hubble constant &$ H_0$ & 70 & 40 $\rightarrow$ 100 \\
Redshift of reionization& $z_{re}$& 11 & 3 $\rightarrow$ 20  \\
Scalar spectral index& $n_s$ & 0.96 & 0.5 $\rightarrow$ 1.3 \\
Normalization & ${\rm ln}[10^{10}A_s]$ & 3.264 & 2.7 $\rightarrow$ 4 \\
Neutrino density & $\Omega_{\nu}h^2$&0.01 & 0 $\rightarrow$ 0.3 \\
Neutrino  degeneracy parameter &$\xi_{\nu}$ & 0 & -1 $\rightarrow$ 1 \\
Number of extra rel. d.o.f.&  $\Delta N_{eff}^{oth}$& 0.046 & -3 $\rightarrow$ 3 \\
Helium mass fraction&$Y_p$ & 0.248 & $\,\, 0.07$ $\rightarrow$ 0.6\\
\hline \hline
\end{tabular}
\end{center}
\end{table}
For the forecast from {\sc Planck}-like simulated data we use the CMB temperature (TT), the
polarization (EE) and their cross-corelation (TE) power spectra  obtained for our fiducial 
cosmological model for multipoles up to $l=2000$ and the expected  experimental characteristics of the {\sc Planck} frequency channels presented in Table~2 \cite{Blue}.  We assume a sky coverage of $f_{sky}=0.8$.\\
Following the method described in Ref.\cite{Perotto,Popa}, for each frequency channel we consider an homogeneous detector noise with the power spectrum given by:
\begin{eqnarray}
N^{c}_{l,\nu}=(\theta_b \Delta_a)^2 \exp^{ l(l+1) \theta^2_b / 8 \ln 2}
\hspace{0.3cm} c \in(T,P) \,
\end{eqnarray}
where $\nu$ is the frequency of the channel, 
$\theta_b$ is the FWHM of the  beam and
$\Delta_c$ are the corresponding sensitivities per pixel of temperature (T) and polarization (P) maps. 
The global noise of the experiment is obtained as:
\begin{equation}
N^c_l=\left[ \sum_{\nu} (N^{c}_{l,\nu})^{-1} \right]^{-1} \, .
\end{equation}
\begin{table}
\begin{center}
\caption{The expected  experimental characteristics
for the {\sc Planck} frequency channels considered in the paper.
$\Delta_T$ and $\Delta_P$ are the sensitivities per pixel for temperature
and polarization maps.We assume a sky coverage of $f_{sky}=0.8$}
\end{center}
\vspace{0.3cm}
\begin{center}
\begin{tabular}{cccc}
\hline \hline
 $\nu $ & FWHM  & $\Delta_T  $ & $\Delta_P $ \\
 (GHz)&(arc-minutes)&    ($\mu$ K)&       ($\mu$ K)                           \\
\hline \hline
               100 & 9.5 &6.8 & 10.9 \\
               143 & 7.1 &6.0 & 11.4 \\
               217 & 5.0& 13.1 & 26.7 \\
\hline \hline
\end{tabular}
\end{center}
\end{table}
We assume uniform prior probability on  parameters $\Theta$ (i.e. we assume that all values of parameters are equally probable) and compute the cumulative distribution function
$C(\theta)=\int^{\Theta}_{\Theta_{min}} {\cal L }
(\Theta)d \Theta/ \int_{\Theta_{min}}^{\Theta_{max}} {\cal L }d \Theta \,$, 
 quoting as upper and lower intervals at 68\% CL 
 the values at which $C(\theta)$ is 0.84 and 0.16 respectively.  
For the case of neutrino mass, $m_{\nu}$,  
we quote  the upper limit at 95\%CL to facilitate the comparison with other measurements.

\section{Results}

We start by making a consistency check, verifying that by using WMAP5+All data
and imposing  $\xi_{\nu}=0$ and $\Delta N_{eff}^{oth}=0$
priors we obtain results in agreement with the ones
obtained by WMAP collaboration \cite{Dunkley,Komatsu}.

In order to understand how the extra relativistic energy density
and the leptonic asymmetry affect the determination of  other
cosmological parameters, we compute first the likelihood functions  for WMAP5+All
by imposing $\xi_{\nu}=0$ prior, extending then  our computation over the whole parameter space for WMAP5+All 
and {\sc Planck}-like simulated data. 
In Figure~1 we compare the marginalized likelihood probabilities obtained for the main cosmological parameters.
\begin{figure}
\begin{center}
\includegraphics[height=10cm,width=10cm]{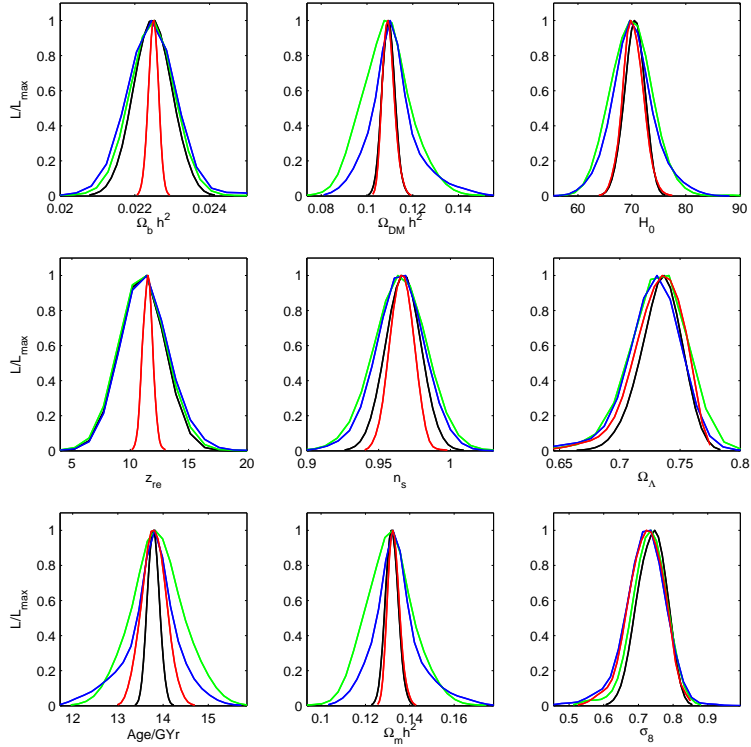}
\caption{The marginalized posterior likelihood probabilities of the main cosmological
parameters obtained for: WMAP5+All with $\Delta N^{oth}_{eff}=0$ and $\xi_{\nu}=0$ priors (black lines), 
WMAP5+All  with $\xi_{\nu}=0$ prior (green lines), WMAP5+All with no neutrino priors (blue lines)
and {\sc Planck}-like simulated data with no neutrino priors (red lines).}
\label{fig1}
\end{center}
\end{figure}

As neutrinos with eV mass decouple when they are still relativistic ($T_{dec} \sim $ 2 MeV),
the main effect of including $\Delta N_{eff}$ is the change of relativistic energy density. 
This changes the redshift of matter-radiation equality, $z_{eq}$, that affects the determination of 
$\Omega_m h^2$ from CMB measurements because of its linear dependence 
on $N_{eff}$ \cite{Komatsu}:
\begin{equation}
1+z_{eq}=\frac{\Omega_m h^2}{\Omega_{\gamma}h^2}\frac{1}{1+0.2271 N_{eff}} \,.
\end{equation}
Here $\Omega_{\gamma}h^2$=2.469 $\times$10$^{-5}$ is the present photon energy density 
parameter for 
$T_{cmb}=~2.725$~K. 
As a consequence, $N_{eff}$ and $\Omega_mh^2$ are linearly correlated, 
with the width of  degeneracy line  given by the uncertainty in the determination of $z_{eq}$.\\
In Figure 2 we compare the marginalized posterior likelihood probabilities of  neutrino parameters,  
$Y_p$ and  $z_{eq}$ obtained in our models. 
The mean values of these parameters and the corresponding (68\% CL) error bars are given in Table 3. \\
\begin{figure}
\begin{center}
\includegraphics[height=7cm,width=10cm]{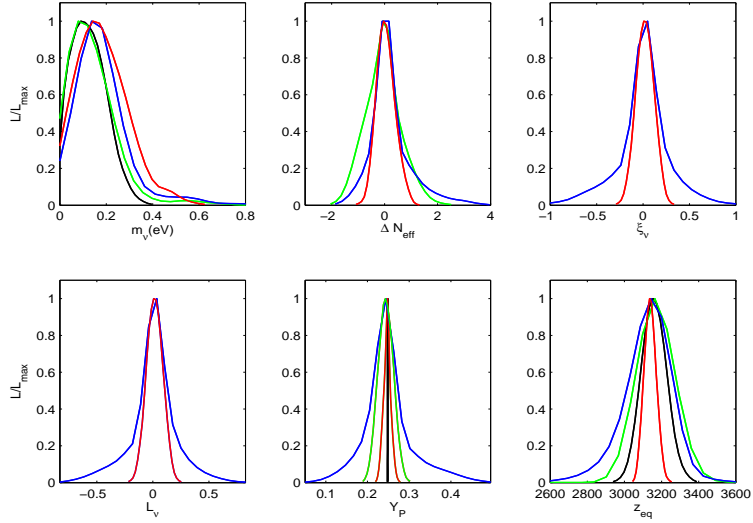}
\caption{The marginalized posterior likelihood probabilities of the neutrino mass, $m_{\nu}$, degeneracy parameter, $\xi_{\nu}$, leptonic asymmetry, $L_{\nu}$, helium mass fraction, $Y_p$, and  redshift of matter-radiation equality, $z_{eq}$, for: 
WMAP5+All with $\Delta N^{oth}_{eff}=0$ and $\xi_{\nu}=0$ priors (black lines), 
WMAP5+All  with $\xi_{\nu}=0$ prior (green lines), WMAP5+All with no neutrino priors (blue lines)
and {\sc Planck}-like simulated data with no neutrino priors (red lines).} 
\label{fig1}
\end{center}
\end{figure}
The LSS measurements provides an independent constraint on $\Omega_mh^2$ 
which helps to reduce the  degeneracy between this parameter and $z_{eq}$. 
From WMAP5+All data with $\Delta N^{oth}_{eff}=0$ and  $\xi_{\nu}=0$ priors we find a mean value of $z_{eq}=3158 \pm 68$ (68\% CL).
One should note that mean value of $z_{eq}$ 
for the standard $\Lambda$CDM model with $N_{eff}=3.046$ is  
$z_{eq}=3176^{+151}_{-150}$ (68\%CL) from WMAP5 data only \cite{Komatsu}.

Figure 3  presents the joint two-dimensional  marginalized distributions 
(68\% and 95\% CL) in $\Omega_mh^2$ - $N_{eff}$ plane (left panel) and $\Omega_mh^2$ - $z_{eq}$ plane (right panel). 
The thick solid  lines in the left panel  show the 68\%  and 95\% CL limits calculated from the corresponding limits on $z_{eq}$ obtained form the WMAP5+All with $\xi_{\nu}=0$ prior by using equation (10). \\
When we transform $N_{eff}$ axis of the left panel to $z_{eq}$ axis from right panel we observe a strong degeneracy between $z_{eq}$ and $\Omega_mh^2$.
This is valid for both WMAP5+All with $\xi_{\nu}$=0 prior and 
WMAP5+All with no neutrino priors.
For the last case, the non-zero neutrino chemical potential augments the value  of $N_{eff}$ by $\Delta N_{eff}(\xi_{\nu})$ as given in equation (1). This imply a larger expansion rate of the Universe, an earlier weak process freeze out with a higher value for the neutron to proton density ratio, and thus a larger value of $Y_p$. On the other hand, a non-zero value of the electron neutrino chemical potential shifts the neutron-proton beta equilibrium, leading to a large variation 
of $Y_p$ \cite{Hamann07,Ichikawa071}.\\
Left panel from Figure 4 presents the  two-dimensional marginalized joint probability distributions (68\% and 95\% CL), showing the degeneracy between $Y_p$ and $N_{eff}$. The total effect of $\xi_{\nu} \neq 0$ is a noticeably increase of the  projected error on $N_{eff}$.

\begin{figure}
\begin{center}
\includegraphics[height=6cm,width=12cm]{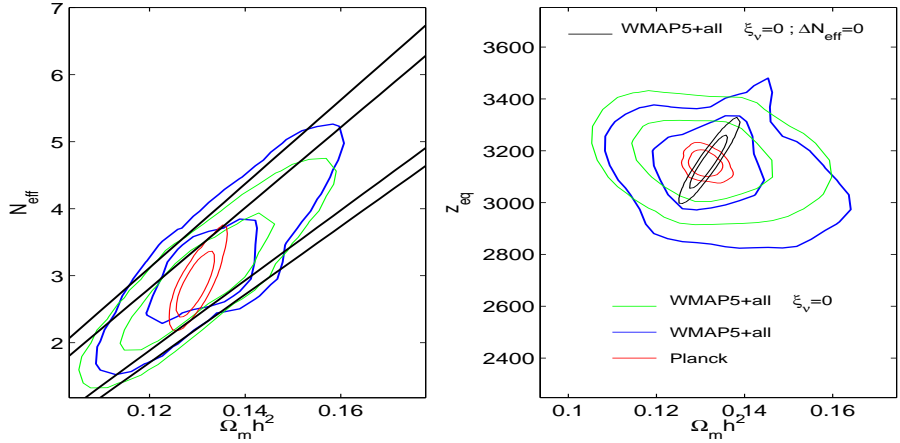} 
\caption{Two-dimensional marginalized joint probability distributions (68\% and 95\% CL) showing  the degeneracy between: $\Omega_mh^2$ and $N_{eff}$ (left panel) and
$\Omega_mh^2$ and $z_{eq}$  (right panel). 
The thick solid  lines in the left panel  show the 68\%  and 95\% CL limits calculated from the corresponding limits on $z_{eq}$ obtained form the WMAP5+All with $\xi_{\nu}=0$ prior by using equation (10). 
The contours show: WMAP5+All with $\Delta N^{oth}_{eff}=0$ and $\xi_{\nu}=0$ priors (black lines),
WMAP5+All with $\xi_{\nu}=0$ prior (green lines), WMAP5+All with no neutrino priors (blue lines) 
and {\sc Planck}-like simulated data with no neutrino priors (red lines).}.
\end{center}
\label{fig1}
\end{figure}
As the anisotropic stress of neutrinos leaves distinct signatures in the CMB 
power spectrum which are not degenerated with $\Omega_mh^2$, we conclude from this analysis  that we observe a non-zero value of $N_{eff}$ from WMAP5+All data
mainly via the change of $z_{eq}$ rather than by the effect of neutrino anisotropic stress. \\ 
Our conclusion is not in agreement with the claim of WMAP team concerning the strong evidence of neutrino anisotropic stress from a similar analysis of the WMAP5+BAO+SN+HST data \cite{Komatsu}. 

We obtain from our analysis a mean value of $N_{eff}=3.026^{+0.638}_{-0.690}$ (68\% CL) from WMAP5+All with $\xi_{\nu}=0$ prior and  $N_{eff}=3.256^{+0.607}_{-0.641}$ (68\% CL) form WMAP5+All with no neutrino priors. 
These values bring an important improvement over the similar 
result obtained from WMAP5+BAO+SN+HST data: $N_{eff}=4.4 \pm 1.5$ (68\% CL).\\
From {\sc Planck}-like simulated data the 68\% error is 
$\sigma (N_{eff}) \approx 0.3$.

The analysis of WMAP5+All data with $\xi_{\nu}=0$ prior provides a strong bound on helium mass fraction of $Y_p=0.2486 \pm 0.0085$ (68\% CL),
that rivals the  bound on $Y_p$ obtained from the conservative analysis of the present data on helium abundance \cite{Olive}.\\
Under the assumption of degenerated BBN this bound is weakened, leading to 
$Y_p=~0.2487^{+0.0451}_{-0.0484}$ (68\% CL) from WMAP5+All with no neutrino priors, that reflects the strong dependence  of  $Y_p$ on $\xi_{\nu}$. \\
From {\sc Planck}-like simulated data the 68\% error on $Y_p$ is 
$\sigma(Y_p)=0.0133$, fully consistent with $Y_p$ bounds obtained 
from the conservative analysis of the present data on helium abundance.
 
We get for neutrino degeneracy parameter a bound of  $-0.216 \leq \xi_{\nu} \leq 0.226$ (68\%CL) that represent an important improvement over the similar result obtained by using the WMAP 3-year data \cite{Lattanzi05}.\\
The CMB only is able to constrain $\xi_{\nu}$ through its contribution to the radiation energy density during radiation domination epoch and the BBN constraints
on $Y_p$.  \\
We find that the sensitivity of CMB to $Y_p$ from {\sc Planck}-like simulated data 
will reduce the error on $\xi_{\nu}$  down to $\sigma(\xi_{\nu}) \simeq 0.089$ (68\% CL), value fully consistent with the BBN bounds.\\
In the right panel from Figure 4  we compare the  two-dimensional marginalized joint probability distributions (68\% and 95\% CL) between $N_{eff}$ and $\xi_{\nu}$, 
showing the potentiality of the future high sensitivity CMB measurements 
to reduce the degeneracy between these parameters.

The inclusion in the analysis of LSS data reduces significantly the upper limit 
of the neutrino mass to $m_{\nu} < 0.284$ eV (95\% CL) from WMAP5+All with 
$\Delta N_{eff}=0$ and $\xi_{\nu}=0$ priors. 
One should note that  the analysis of WMAP5+BAO+SN+HST leaded to $m_{\nu} < 0.61$ eV (95\% CL) \cite{Komatsu}.\\ 
Under the assumption of degenerated BBN this bound is weakened due to the 
degeneracy between $N_{eff}$ and $\xi_{\nu}$. From WMAP5+All with no neutrino priors 
we find $m_{\nu} < 0.535$ eV (95\%CL), while from {\sc Planck}-like simulated data 
with no neutrino priors we obtain $m_{\nu} < 0.440$ eV. 
\begin{figure}
\begin{center}
\includegraphics[height=6.cm,width=12.cm]{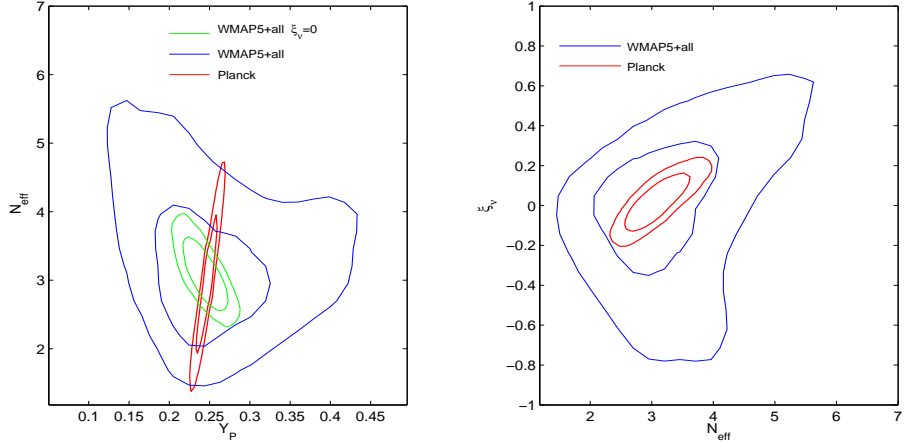} 
\caption{ Two-dimensional marginalized joint probability distributions (68\% and 95\% CL) showing the degeneracy between: $Y_p$ and $N_{eff}$ (left panel) and
$N_{eff}$ and $\xi_{\nu}$  (right panel). 
The contours show: WMAP5+All with $\xi_{\nu}=0$ prior (green lines), WMAP5+All with no neutrino priors (blue lines) 
and {\sc Planck}-like simulated data with no neutrino priors (red lines).}
\label{fig1}
\end{center}
\end{figure}

\section{Summary and conclusions}

In this paper, we set bounds on the radiation content of the Universe and neutrino properties by using the WMAP-5 year CMB measurements complemented  with 
most of the existing CMB  and  LSS measurements (WMAP5+All),
imposing also self-consistent BBN constraints on the primordial helium abundance, which proved to be important in the estimation of cosmological parameters \cite{Hamann07, Ichikawa071}.

We consider lepton asymmetric cosmological models parametrized by the neutrino degeneracy parameter $\xi_{\nu}$ and the variation of the relativistic degrees of freedom, $\Delta  N^{oth}_{eff}$, due to possible other physical processes  occurred between BBN and  structure formation epochs.

From our analysis we get a mean value of the effective number of relativistic neutrino 
species form WMAP5+All with no neutrino priors, of $N_{eff}=3.256^{+0.607}_{-0.641}$ (68\% CL),  bringing an important improvement over the similar result obtained from WMAP5+BAO+SN+HST data \cite{Komatsu}.\\
Although the LSS measurements provides an independent constraint on $\Omega_mh^2$ 
which helps to reduce the  degeneracy between this parameter and the redshift of matter radiation equality, $z_{eq}$, we find a strong correlation between $\Omega_mh^2$ and $z_{eq}$, showing that we observe a non-zero $N_{eff}$ value from WMAP5+All data mainly due to the change of $z_{eq}$, rather than by the neutrino anisotropic stress. 

The analysis of WMAP5+All data with $\xi_{\nu}=0$ prior provides a strong bound on helium mass fraction of $Y_p=0.2486 \pm 0.0085$ (68\% CL),
that rivals the  bound on $Y_p$ obtained from the conservative analysis of the present data on helium abundance \cite{Olive}.
Under the assumption of degenerated BBN this bound is weakened to 
$Y_p=0.2487^{+0.0451}_{-0.0484}$ (68\% CL), reflecting the strong dependence of  $Y_p$ on $\xi_{\nu}$. 

From the analysis of WMAP5+All data we get for neutrino degeneracy parameter a bound of  $-0.216 \leq \xi_{\nu} \leq 0.226$ (68\%CL).

The inclusion in the analysis of LSS data reduces  the upper limit 
of the neutrino mass to $m_{\nu} < 0.535$ eV (95\% CL), from WMAP5+All with no neutrino priors, with respect to the values obtained from the analysis from WMAP5-only data \cite{Dunkley} and WMAP5+BAO+SN+HST data \cite{Komatsu}.

The analysis of WMAP5+All measurements bring also an important improvement over the similar results obtained by using WMAP~1-year measurements complemented with LSS data \cite{Crotty03} and the WMAP 3-year data alone \cite{Lattanzi05,Ichi08}.

We forecast that the CMB temperature and polarization measurements observed 
with high angular resolutions and sensitivity by the future {\sc Planck} satellite will reduces the errors on $\xi_{\nu}$ and $Y_p$ down 
to $\sigma(\xi_{\nu}) \simeq 0.089$ (68\% CL) and 
$\sigma(Y_p)=0.0133$ (68\% CL) respectively,  
values fully consistent with the BBN bounds on these parameters \cite{Rafelt}.\\
Our forecasted errors on the cosmological 
parameters from {\sc Planck}-like simulated data are 
also consistent those obtained in Ref.\cite{Hamann07}. 

\begin{table}
\begin{center}
\caption{The table shows the mean values and the absolute errors on the main cosmological parameters obtained from the analysis of WMAP5+All data and {\sc Planck}-like simulated data. For all parameters, except $m_{\nu}$, we quote the errors at 68\% CL. For $m_{\nu}$ we give the values of the 95\% upper limit.}
\vspace{0.3cm}
\begin{tabular}{lllll}
\hline \hline   
&Priors: $\xi_{\nu}=0$; $\Delta N_{eff}$=0  & Priors: $\xi_{\nu}=0$ & No neutrino priors &No neutrino priors \\
&WMAP5+All  & WMAP5+All & WMAP5+All& {\sc Planck} \\
Parameter &           &                  &          &            \\
\hline  
$\Omega_{b}h^2$ &$0.02247 \pm 0.00053$ &$0.02245\pm 0.00059$&$ 0.02246^{+0.00063}_{-0.00072}$&$0.02247 \pm 0.00021 $ \\
$\Omega_{dm}h^2$&$0.1094 \pm 0.0028$   &$ 0.1086\pm  0.0078$&$0.1115^{+0.0089}_{-0.0093}$&$0.1110\pm 0.0038$ \\
$ H_0$ &$70.5 \pm 1.7$ &$69.9\pm 3.4$&$70.2^{+3.6}_{-3.8}$&$70.1\pm 2.3$   \\
$z_{re}$&$11.02 \pm 1.71$&$11.07\pm 2.01$&$11.31 \pm ^{1.92}_{2.11}$&$11.01\pm 0.39$   \\
$n_s$ &$0.966 \pm 0.013$ &$0.965\pm 0.019$&$0.965\pm  0.018$&$0.961 \pm 0.008 $ \\
${\rm ln}[10^{10}A_s]$ &$3.264 \pm 0.036$&$3.264 \pm 0.051$&$3.265 \pm 0.056$&$3.264 \pm 0.017$  \\
$m_{\nu}$(eV) & $\le 0.284$&$\le 0.419$& $\le 0.535$& $\le 0.440$\\
$\xi_{\nu}$ & - &-&$0.005 \pm 0.221$&$0.021 \pm 0.089$ \\
${\cal L}_{\nu}$& - &-&$0.001 \pm 0.165$&$0.015 \pm 0.073$ \\
$\Delta N_{eff}$& 0.046 &$ -0.020^{+0.638}_{-0.690}$&$ 0.256^{+0.607}_{-0.641}$&$0.089 \pm 0.288$  \\
$Y_p$ &$ 0.2480 \pm 0.0002$&$0.2486 \pm 0.0085$&$ 0.2487^{+0.0451}_{-0.0484}$&$0.2477\pm 0.0133$ \\
$z_{eq}$&$3158 \pm 68$&$3167^{+103}_{-109}$&$3124^{+120}_{-128}$&$3145 \pm 34$ \\
\hline \hline
\end{tabular}
\end{center}
\end{table}

\vspace{0.3cm}
{\bf Acknowledgements}

We acknowledge the referee for useful comments and suggestions.
L.P. and A.V. acknowledge the support by the ESA/PECS Contract "Scientific exploitation of Planck-LFI data"

We also acknowledge the use of the GRID computing system facility at the Institute for
Space Sciences Bucharest and would like to thank the staff working there.

\section*{References}




q



\end{document}